
\input harvmac

\def \> {\rangle}
\def \< {\langle}

\vskip 0.8cm

\Title{ \vbox{\baselineskip12pt\hbox{  Brown Het-1293}}}
{\vbox{ \centerline{  Higher Dimensional Geometries }
\vskip.06in
\centerline{  from  }
\vskip.06in
\centerline{  Matrix Brane Constructions  } } }

\centerline{$\quad$ {Pei-Ming Ho$^1$, Sanjaye Ramgoolam$^2$}}
\smallskip
\centerline{{\sl $^1$ Department of Physics}}
\centerline{{\sl National Taiwan University}}
\centerline{{\sl Taipei 106, Taiwan, R.O.C.}}
\smallskip
\centerline{{\sl $^2$ Department of Physics}}
\centerline{{\sl Brown  University}}
\centerline{{\sl Providence, RI 02912 }}
\smallskip
\centerline{{\tt pmho@phys.ntu.edu.tw, ramgosk@het.brown.edu}}

 \vskip .3in 

  Matrix descriptions of even dimensional 
  fuzzy spherical branes $S^{2k} $ in Matrix Theory and other 
  contexts in Type II superstring theory 
  reveal, in the large $N$ limit, higher dimensional 
  geometries  $SO(2k+1)/U(k)$,  
  which have  an interesting spectrum of $SO(2k+1)$ harmonics 
 and can be  up to $20$ dimensional, while the spheres are 
  restricted to be of dimension less than $10$.
  In the case $k=2$, the matrix description   has two dual
   field theory formulations.  
  One involves  a field theory living on the non-commutative  
 coset $SO(5)/U(2)$   which is a fuzzy $S^2$ fibre 
 bundle over a fuzzy $S^4$. In the other, 
  there is a $U(n)$ gauge theory on a fuzzy $S^4$ 
  with $ { \cal O }( n^3)$ instantons. 
  The two descriptions can be related by exploiting the 
  usual relation between the fuzzy  two-sphere and $U(n)$ Lie algebra. 
  We discuss the analogous phenomena in the higher dimensional cases, 
  developing a relation between fuzzy $SO(2k)/U(k)$
  cosets and  unitary Lie algebras.

\def\Ga{ \Gamma } 

\def\can{ {\cal{A}}_n }   
\def\chan{ \hat {\cal{ A }}_n }
\def\co{ {\cal{O}} }

\Date{ {\it Nov  2001} } 

\lref\gkp{ H. Grosse, C. Klimcik and P. Presnajder,
``On finite  4-D Quantum Field Theory in noncommutative geometry,''
[hep-th/9602115],  Commun.Math.Phys.180:429-438,1996.
}

\lref\fzalg{ S.~Ramgoolam,
``On spherical harmonics for fuzzy spheres in diverse dimensions,''
Nucl.\ Phys.\ B {\bf 610}, 461 (2001)
[hep-th/0105006].
}

\lref\clt{J. Castellino, S. Lee, W. Taylor,
``Longitudinal 5-branes as 4-spheres in matrix theory,'' 
[hep-th/9712105], Nucl.Phys. B526 (1998) 334-350.
} 

\lref\cmt{ N. Constable, R. Myers and O. Tafjord,
``Non-abelian brane
intersections,''
[hep-th/0102080],  JHEP 0106:023,2001.
}

\lref\kramer{ M. Kramer, `` Some remarks suggesting an interesting 
 theory of harmonic functions on $SU(2n+1)/Sp(n)$ and 
$SO(2n+1)/U(n)$,'' Arch. Math. 33 ( 1979/80), 76-79. }

\lref\iso{ S.~Iso, Y.~Kimura, K.~Tanaka and K.~Wakatsuki,
``Noncommutative gauge theory on fuzzy sphere from matrix model,''
Nucl.\ Phys.\ B {\bf 604}, 121 (2001)
[hep-th/0101102].
}

\lref\gelf{ I.M.Gelfand, A.V. Zelevinskii,
``Models of representations
of classical groups and their hidden symmetries,'' 
Funct. An. Priloz. vol. 18, 1984. } 

\lref\zr{ Z. Guralnik,  S. Ramgoolam, 
``On the polarization of unstable D0-branes
 into non-commutative odd spheres,''
[hep-th/0101001], JHEP 0102 (2001) 032.
}

\lref\ho{ P.~M.~Ho,
``Fuzzy sphere from matrix model,''
[hep-th/0010165], JHEP {\bf 0012}, 015 (2000).
}
\lref\antram{ A. Jevicki and S. Ramgoolam,
`` Noncommutative gravity from the ADS/CFT correspondence,''
[hep-th/9902059], JHEP {\bf 9904} (1999) 032.
 }

\lref\holi{
P.~M.~Ho and M.~Li,
``Fuzzy spheres in AdS/CFT correspondence
and holography from  noncommutativity,''
Nucl.\ Phys.\ B {\bf 596}, 259 (2001)
[hep-th/0004072].
P.-M.~Ho and M.~Li,
``Large N expansion from fuzzy AdS(2),''
Nucl.\ Phys.\ B {\bf 590}, 198 (2000)
[hep-th/0005268].
}
\lref\gutze{F. Gursey and C.H. Tze, ``On the role of division,
Jordan and related algebras in particle physics,''
World Scientific 1996.}

\lref\wang{H. C. Wang,
Nagoya. Math. J. {\bf 13}, 1 (1958).}

\lref\Myers{ R. C. Myers,
``Dielectric-branes'',JHEP 9912 (1999) 022
[hep-th/9910053].
}

\lref\fulhar{ W. Fulton and G. Harris,
``Representation theory,'' 
Springer Verlag 1991.} 

\lref\BFSS{ ``T. Banks, N. Seiberg and S. Shenker, 
M-Theory as a matrix model : a conjecture,''
[hep-th/9610043],  Phys.Rev.D55:5112-5128,1997.
 } 

\lref\hrt{ P.M.Ho, S.Ramgoolam and R.Tatar,
``Quantum space-times and finite N
effects in 4-D superyang-mills theories,''
[hep-th/9907145], 
Nucl.Phys.{\bf B573} (2000) 364,
}

\lref\dwhn{  B. de Wit, J. Hoppe, H. Nicolai, 
``On the quantum mechanics of supermembranes,''
Nucl.Phys.B305:545,1988.
}

\lref\madore{ J. Madore, ``The Fuzzy sphere, ''
Class.Quant.Grav.9:69-88,1992  
}

\lref\bv{ M. Berkooz and H. Verlinde,
``Matrix theory, AdS/CFT and
Higgs-Coulomb equivalence,''
[hep-th/9907100], 
JHEP 9911 (1999) 037.
} 

\lref\bss{ T. Banks, N. Seiberg, S. Shenker, `` Branes form
Matrices,'' [hep-th/9612157], \hfill\break
Nucl.Phys. B490:91-106,1997 
}

\lref\grt{ O. Ganor, S. Ramgoolam, W. Taylor,
``Branes, fluxes and duality in M(atrix)-theory,''
[hep-th/9611202], 
 Nucl.Phys.B492:191-204,1997 
}

\lref\AH{ N.~Arkani-Hamed, A.~G.~Cohen and H.~Georgi,
``(De)constructing dimensions,''
[hep-th/0104005],
Phys.\ Rev.\ Lett.\  {\bf 86}, 4757 (2001).
}

\lref\bdlmo{ 
A.P. Balachandran, B. P. Dolan, J.H. Lee, X. Martin, D. O'Connor
``Fuzzy complex projective spaces and their star products,''
[hep-th/0107099]
 }

\lref\tv{
S. P. Trivedi, S. Vaidya,
``Fuzzy cosets and their gravity duals,''
[hep-th/0007011], JHEP 0009:041,2000 
}

\lref\connes{ A. Connes, ``Non-commutative geometry,'' Academic Press,
 1994.  }   

\lref\ps{ P.J.Silva,`` Matrix string theory and the Myers effect,''
[hep-th/0111121].
}

\lref\rief{ M.A.Rieffel
``Matrix algebras converge to the sphere 
for quantum Gromov--Hausdorff distance,'' [math.OA/0108005].
}

\lref\nara{V.P. Nair, S. Randjbar-Daemi,
``On brane solutions in M(atrix) theory,'' Nucl.Phys. B533 (1998)
333-347 
}

\newsec{ Introduction } 

 Fuzzy spheres of dimension higher than two
have found a variety of physical applications
\gkp\clt\ho\cmt\zr\  and may be relevant to 
the ADS/CFT correspondence \antram\hrt\bv\holi. 
In this paper we will study some 
geometrical aspects of the matrix algebras which 
are related to these higher fuzzy spheres. 
 This will allow us to observe some interesting 
 features associated with the physics of fluctuations 
 around  solutions of brane actions involving these fuzzy 
 spheres. 

 In section two we review some background on 
 higher dimensional fuzzy spheres. 
 Fuzzy spheres are described by looking for an
$ N \times N $  matrix realization of the equation 
$$ \sum_{\mu =1 }^{2k+1} X_{\mu} X_{\mu} = 1, $$
 which classically describes a sphere. The $X_{\mu}$
matrices transform in the vector representation of
 $SO(2k+1)$ symmetry group of the sphere.  
The decomposition of the matrix algebras into 
 irreducible representations of 
 $SO(2k+1)$ is reviewed. It has been pointed out 
 that only a subset of these representations 
 approaches the space of spherical harmonics on a 
 classical sphere \fzalg. This raises the issue of the 
 geometry of the large $N$ limit of the full matrix algebra. 
 Some relevant mathematical results in harmonic 
 analysis are noted,  which indicate  that the relevant space is 
 $SO(2k+1)/U(k)$.

 In section three we describe some aspects of the system 
 of algebraic equations obeyed by the 
 generators of the matrix algebra. 
 The $X_{\mu}$'s are among the 
 generators and they are obtained from the action 
 of gamma matrices $\Gamma_{\mu}$ of $SO(2k+1)$ 
 on $ Sym ( V^{\otimes n } )$, an irreducible representation 
of $ SO(2k+1)$ obtained by symmetrizing the $n$-fold tensor power
 of the fundamental spinor. This representation is $N$ dimensional 
where $ N \sim n^{ k(k+1)} $,  and the $X^{\mu}$ are $ N \times N $ matrices.
The matrix algebra is denoted as $\chan ( S^{2k} ) $. 
 While the $X_{\mu}$  do generate $\chan ( S^{2k} ) $
 under matrix multiplication at finite $n$, 
consideration of the large $n$ limit indicates that it is useful 
to introduce extra generators transforming in antisymmetric 
 tensor representations of $SO(2k+1)$, and obeying some constraint
 equations. 
We describe further geometrical structures  on  $\chan ( S^{2k} ) $,
namely an algebra of derivations $SO(2k+1,1)$. 
It is useful to distinguish the matrix algebra $ \chan ( S^{2k } ) $
from a subspace $ \can ( S^{2k} )$ which decomposes 
into symmetric traceless representations of $SO(2k+1)$. 
This subspace $ \can ( S^{2k} ) $ has a non-associative 
multiplication which becomes associative in the large $N$ 
limit \fzalg. It is shown that $SO(2k+1) $ acts as 
 derivations of this non-associative algebra. 
Further properties of  $ \can ( S^{2k} ) $ are developed 
in the Appendix.

In section four we consider, in detail, the 
fuzzy four-sphere. We show that the 
large $n$ limit of $ \chan( S^4)$ describes the algebra of functions 
on $SO(5)/U(2)$. This is done by considering 
 a special solution to the system of equations 
obeyed by the generators of the matrix algebra
 in the large $N$ limit. Rather than writing out all 
 the equations and looking for solutions, we 
 use the simple fact that the explicit matrix construction 
of the generators should give solutions to the equations. 
 Following  the relation between operators and classical 
 variables in quantum mechanics, we obtain these solutions  
by considering the expectation values of these generators in a 
state. Given the solution, we find its stabilizer group 
 to be $U(2)$. Since the system of equations obeyed by the 
 generators is $SO(5)$ covariant, the special solution 
 generates $SO(5)/U(2)$. The harmonics on this space 
 coincide with the large $N$ decomposition of the matrix algebra, 
thus showing that the matrix algebra , $ \chan ( S^{4} ) $,
  approaches the space of functions on this coset. 

The coset $SO(5)/U(2)$ is a bundle over $S^4$ with 
fibre which is a two-sphere. We describe a finite 
$n$ version of this bundle structure, with the fibre 
 being a fuzzy two-sphere. The connection between the fuzzy 
two-sphere and $U(n+1)$ is exploited to show that the 
antisymmetric generators $X_{i j} $ which 
generate the algebra of functions on the fibre 
 can be associated with field strengths of a $U(n+1)$
 bundle at a point on the $S^4$. In this way, we 
 establish a direct relation, by construction so to speak, 
between the matrix description of the worldvolume 
theory of $N \sim n^3/6 $ $D0$ branes
  with the field strengths of the dual instanton description in terms 
of $U(n+1)$ gauge theory with $N \sim n^3/6$ instantons
on the worldvolume of $n+1$ spherical $4$-branes, a connection 
 expected on general physical grounds \clt\cmt. 

In section five and six we develop 
analogous points for the fuzzy $S^6$ and 
the fuzzy $S^8$. This uses an extension of the 
connection between unitary Lie algebras and fuzzy two-spheres 
\dwhn\ to connections between unitary Lie algebras 
and $SO(2k)/U(k)$ for $k=3,4$. This is expected to generalize 
to any $k$. 

In section seven, we observe that fluctuations of 
branes described in the fuzzy sphere constructions
have in addition to a description in terms of non-Abelian gauge theory
 on spheres with a large number of instantons, 
a description in terms of a field theory on the $SO(2k+1)/U(k)$ coset.
The field theory has symmetries which include a non-commutative 
$U(1)$ gauge symmetry 
and a global $SO(2k+1)$ symmetry. A complete understanding of the
higher dimensional action is left for the future.

 In the Appendix we show  that the non-associative algebra
$ \can (S^{2k} ) $
is actually commutative at finite $n$. If all $ \co ( 1/n ) $ 
terms are dropped from the structure constants, the algebra 
is commutative and associative and coincides with the 
 algebra of spherical harmonics. If  $ \co ( 1/n ) $ 
terms are kept, and $  \co ( 1/n^2 ) $ terms are dropped, 
then the algebra is an infinite dimensional Jordan Algebra. 
At higher orders in $ 1/n $ and at finite $n$, it is a
 more general commutative and non-associative algebra which is not
Jordan. For concreteness these points are developed for the case 
$k=2$ but the arguments are of a general nature.

\newsec{ Review of relevant background } 

Here we will review some material 
from  physics and mathematics 
literature which will be useful. 

\subsec{ Fuzzy spheres and matrix brane constructions }

The  fuzzy four-sphere  was used  in the 
construction of time-dependent 4-brane solutions 
from the action of zero-branes in \clt, 
in the context of Matrix Theory of BFSS.  
On the other hand,
the Lagrangian for a matrix model coupled to totally
antisymmetric tensor field of rank $(2k+1)$ is of the form
\Myers\
$$ L=Tr\left({1\over 2}[\Phi_{\mu},\Phi_{\nu}]^2
+F\epsilon_{\mu_1\cdots\mu_{2k+1}}
\Phi_{\mu_1}\cdots\Phi_{\mu_{2k+1}}\right), $$
where $F$ is the field strength.
An example of this was found by Myers \Myers\
that D0-branes expand into spherical branes
in the background of RR gauge fields.
Along similar lines, one considers 
the worldvolume action for D-strings 
 ending on D5-branes and in this case
 $5$ matrices transverse to the D-string worldvolume 
 describe the end-point acquiring a spherical 
4-dimensional geometry\cmt.  A discussions of fuzzy 
spheres in Matrix theory also appears in \nara. 
A different application involves the study of
compactifications of M-theory in the context of
matrix model compactified on
a $(2k+1)$-dimensional sphere.
The Lagrangian is of the form
$$ L=Tr\left({1\over 2}[\Phi_{\mu}, \Phi_{\nu}]^2
-a(\Phi_{\mu}\Phi_{\mu}-R^2)\right), $$
where $a$ is the Lagrange multiplier. 
The $X_{\mu} $ which are used to construct 
fuzzy sphere coordinates in the next section
can be used to solve the equations of motion
in all of the above cases by setting
$\Phi_{\mu}=L X_{\mu}$
for some constant $L$.

Since the algebra of functions of $X$
is equivalent to the algebra of $N\times N$ matrices,
in these cases, generic configurations of the matrix variables
can be described as fields 
living on a manifold whose algebra of functions 
is approximated by the large $N$ limit of $\chan( S^{2k} )$.  
In this paper, we will describe this 
 manifold, and relate its finite $n$ 
non-commutative geometry to non-Abelian 
gauge theory on the fuzzy sphere. The rank and instanton 
number of the gauge bundle are both related to 
$N$ which sets the scale of non-commutativity, 
so these constructions are far from simply 
 being discretized versions of spheres.


\subsec{ Decomposing matrix algebras into reps. of $SO(2k+1)$ } 

 The construction of fuzzy spheres $S^{2k}$ 
 starts with the action of $X_{\mu}$, $ \mu =1 \cdots 2k+1$, 
 which are obtained from an action of $\Gamma_{\mu}$ 
 on $Sym ( V^{\otimes n } ) $:
\eqn\actx{ X_{\mu} = \sum_{r=1}^{n} \rho_{r} ( \Gamma_{\mu} ). } 
 The operator $ \rho_r( \Gamma_{\mu} ) $ acts on 
 the $r$-th copy of $V$ in $ V^{\otimes n } $. 
 Since the sum \actx\ is symmetric it gives a well 
 defined action on     $Sym ( V^{\otimes n } ) $.
Each $X_{\mu} $ is an $N \times N $ matrix where 
 $N \sim n^{k(k+1) } $ at large $n$. 
 Products of the $X_{\mu}$ generate the full set of 
matrices. Different products can be used to generate 
sets of matrices transforming according to different 
 representations of $SO(2k+1)$. Each tensor representation 
occurs once, with a cutoff on the first row of the Young diagram 
 corresponding to the representation of $r_1 \le n$. 
With this cutoff, the dimensions of these representations 
add up to $N^2$, exactly the size of the matrix algebra \fzalg. 
In the large $n$ limit, all representations of $SO(2k+1)$ occur 
with unit multiplicity. 

The algebra of functions on the sphere $S^{2k}$ 
decomposes into representations which are symmetric and traceless, 
i.e they correspond to Young diagrams with row lengths $r_i =0$ 
for $i > 1$. To get this algebra we have to project 
out the representations which have any columns of length greater 
than $1$. On this projected space we have a non-associative 
multiplication.  

In this paper we will be focusing on the geometry 
of the full matrix algebra. Since the matrix algebra 
has a simple decomposition into representations
 of $SO(2k+1)$, this can be expected to give some strong 
hints about the geometry related to it. 

\subsec{  A heuristic guess } 

 The operators $X_{\mu}$ have eigenvalues
 ranging from $-n$ to $n$. The number of degrees
 of freedom in the matrix algebra grows like $N^2$ 
 which is $ n^{k(k+1)}$. If we think of the 
 space described by the full matrix algebra 
 as a discretized space and associate one degree 
 of freedom with each point, we expect the 
 space to be $k(k+1) $ dimensional. This will indeed 
 be the dimension of the coset space $SO(2k+1)/U(k)$ 
 which we will prove to have an algebra of functions 
which is the classical limit 
 of the matrix algebra.

\subsec{ The results of Kramer }

A paper of Kramer \kramer\ has done the job of finding the 
space whose algebra of functions contains 
each representation of $SO(2k+1)$ once. 
It is the space $ SO(2k+1)/U(k)$. This space 
has dimension $ k(k+1)$.
The construction of algebras which contain each representation
of $SO(2k+1)$ is not in fact unique. Such constructions 
are referred to as model spaces, and they can also 
come up as spaces of sections of certain bundles over cosets \gelf.  
To establish that our matrix algebra is indeed associated with 
 $ SO(2k+1)/U(k)$ we will describe the system of equations 
obeyed by these matrices in the large  $n$ limit. 

\newsec{ Description in terms of algebraic equations } 
\subsec{ Explicit form of gamma matrices } 

To fix notation we will give the form of the 
gamma matrices $\Gamma_{\mu} $ with $\mu=1\cdots 2k+1$. 
The gamma matrices of $SO(2k+1)$  obey the equations :  
\eqn\gam{ 
\{ \Gamma_{\mu} , \Gamma_{\nu} \} = 2\delta_{ \mu \nu }. }
They can be expressed 
in terms of a set of fermionic  oscillators $a_{i}$ with 
$i$ running from $1$ to $k$, and obeying 
\eqn\fermdef{ \{ a_{i}, a_{j}^{\dagger} \} = \delta_{ij}. }
The expressions for the gamma matrices are :  
\eqn\gamrep{\eqalign{  
& \Gamma_{2i-1} =  ( a_{i} + a_{i}^{\dagger} ),  \cr 
& \Gamma_{2i} = i ( a_{i} - a_{i}^{\dagger}   ) \cr }}
 for $ i = 1, \cdots, k $  and   
 $  \Gamma_{2k+1} = \Gamma_{1} \cdots \Gamma_{2k} $. 
 
 A $2^k$ dimensional representation of the 
 $ \Gamma $ matrix algebra is obtained by defining 
 a state $ |0\> $ which is annihilated by the 
 fermionic annihilation operators $a_{i}$ and
 acting with the creation operators to generate  
 $2^{k} $ different states.  


\subsec{
Algebra of the fuzzy sphere }

 From the matrices describing the fuzzy spheres, we 
  can write down a set of algebraic equations 
  which are satisfied by the variables. In the large $n$ limit 
  all the variables involved are commutative. 
 
For a fuzzy sphere define $X_{\mu}$ as in \actx. 
Let us denote their commutators by
\eqn\xmn{ X_{\mu \nu } = {1 \over 2} [X_{\mu}, X_{\nu}]
= {1\over 2 }\sum_{r} \rho_r ( \Gamma_{\mu} \Gamma_{\nu} -
\Gamma_{\nu} \Gamma_{\mu} ).
}
For the fuzzy $6$-sphere we also need the totally antisymmetrized
cubic products of gamma matrices: 
\foot{
They are not antisymmetric cubic products of $X_{\mu}$'s,
but are more complicated functions of them.
}
\eqn\xmnl{\eqalign{ 
& X_{\mu \nu \lambda } =  
\sum_{r} \rho_r ( \Gamma_{[\mu} \Gamma_{\nu} \Gamma_{\lambda]} ) \cr 
& = \sum_{r} \rho_r ( \Gamma_{\mu} \Gamma_{\nu} \Gamma_{\lambda} +
\Gamma_{\lambda} \Gamma_{\mu} \Gamma_{\nu} +
\Gamma_{\nu} \Gamma_{\lambda} \Gamma_{\mu } - 
\Gamma_{\nu   } \Gamma_{\mu} \Gamma_{\lambda} - 
\Gamma_{\lambda   } \Gamma_{\nu} \Gamma_{\mu} - 
\Gamma_{\mu   } \Gamma_{\lambda} \Gamma_{\nu}  ) \cr  } }
and similarly, for the fuzzy-8-sphere, we will also need 
\eqn\xmnlr{  X_{\mu \nu \lambda \rho } =  
\sum_{r} \rho_r ( \Gamma_{ [\mu} \Gamma_{\nu} \Gamma_{\lambda}
\Gamma_{\rho ]} ). } 

These matrices satisfy the following algebraic relations:
\eqn\XXa{ X_{\mu}X_{\mu}=c, }
\eqn\XXb{ [X_{\mu\nu}, X_{\rho}]=
2(\delta_{\nu\rho}X_{\mu}-\delta_{\mu\rho}X_{\nu}), }
\eqn\XXc{ [X_{\mu\nu}, X_{\kappa\rho}]=
2(\delta_{\nu\kappa}X_{\mu\rho}+\delta_{\mu\rho}X_{\nu\kappa}
-\delta_{\mu\kappa}X_{\nu\rho}-\delta_{\nu\rho}X_{\mu\kappa}), }
where
\eqn\c{ c=n(n+2k). }
The $X_{\mu\nu}$'s can be identified with
generators of the $SO(2k+1)$ Lie algebra.
Together with $X_{\mu}$,
they generate the Lie algebra of $SO(2k+1,1)$.

In addition, we have generalized self-duality relations
\eqn\SD{ \epsilon_{\mu_1\cdots\mu_{2k+1}}
X_{\mu_1}\cdots X_{\mu_{2k}} = \lambda X_{\mu_{2k+1}}. }
This relation is valid only when
they are viewed as operators on $Sym(V^{\otimes n})$.
For $k=1$, 
$\lambda=2$.
For $k=2$, 
$\lambda=8(n+2)$.
For $k=3$,
$\lambda=144n(n^2+4)$.

To describe the geometry corresponding to this algebra, 
we define variables $ Z_{\mu}$, $Z_{\mu \nu} $, 
$Z_{\mu\nu \lambda}$, $ Z_{\mu \nu \lambda \rho}$, 
which satisfy algebraic equations with real coefficients 
and they generate an algebra which defines a non-commutative  manifold.
They are related to $X$ through the following relations
\eqn\defzx{\eqalign{&   Z_{\mu } \sim  { 1 \over n }X_{\mu}, \cr 
              & Z_{\mu \nu} \sim { i \over n } X_{\mu \nu }, \cr 
 & Z_{\mu \nu \lambda} \sim  { 1 \over n }X_{\mu \nu \lambda}, \cr 
 & Z_{\mu \nu \lambda \rho}
                 \sim { i  \over n }X_{\mu \nu \lambda \rho }. \cr }}
The $Z$'s are suitably normalized so that
they obey equations having coefficients 
$ \co ( 1 )$ in the large $n$ limit. 
In this paper, we will use $Z$ as generators of the 
algebra of functions $ \chan ( S^4) $
and use $X$'s as derivatives.

\subsec{ Derivations on the matrix algebra $ \chan ( S^{2k}  ) $  } 

In the large $n$
limit the matrix algebra reduces to an algebra of 
polynomials in $Z$'s with constraints,
which will be given later.
The action of the operators $X_{\mu} $, $X_{\mu\nu}$
by commutators transforms the polynomials in the variables 
$Z_{\mu } , Z_{\mu \nu} $. 
In particular, the $X_{\mu\nu}$'s satisfy an $SO(2k+1)$ algebra,
organizing all polynomials of $Z$'s into
representations of $SO(2k+1)$.

Recall that we chose the normalizations \defzx\ .
We know from \xmn\ , \XXb\ that 
\eqn\actX{\eqalign{&   [ X_{\mu } , Z_{\nu } ] = - 2i Z_{\mu \nu },  \cr 
        &   [ X_{\mu } , Z_{\nu \lambda} ] = - 2i
( \delta_{\mu \nu } Z_{ \lambda}  - \delta_{\mu \lambda } Z_{\nu }  ). \cr }}
To reproduce these formulae we write : 
\eqn\xform{ ad ( X_{\mu }  )  
= -2i \left( Z_{\mu \nu } { \partial \over \partial Z_{\nu } } + 
 Z_{\nu } { \partial \over \partial Z_{\mu \nu } }\right).} 
Similarly, we know from \XXb\ , \XXc\ that
\eqn\actxij{\eqalign{ 
[ X_{\mu \nu }, Z_{\lambda} ] &= 2(\delta_{\nu \lambda} Z_{\mu}  -
\delta_{\mu \lambda} Z_{\nu }),  \cr 
[ X_{\mu \nu } , Z_{\lambda \rho} ] &= 2(
\delta_{\nu \lambda } Z_{ \mu \rho } -
\delta_{nu \rho} Z_{\mu \lambda} 
 + \delta_{\mu \rho} Z_{\nu \lambda} -
\delta_{\mu \lambda} Z_{\nu \rho}).
 \cr }}
This can be reproduced by 
\eqn\formXij{ ad(X_{\mu \nu}) = 2\left( Z_{\mu }
{ \partial \over \partial Z_{\nu } } - 
 Z_{\nu } { \partial \over \partial Z_{ \mu  } }
 - Z_{\mu \lambda}  {\partial \over \partial Z_{ \lambda \nu } } + 
 Z_{\nu \lambda }  {\partial \over \partial Z_{\lambda \mu } }
\right). } 

For consistency, the equations \xform\ and  \formXij\ 
should agree with 
$[X_{\mu } , X_{\nu } ] = 2X_{\mu \nu } $, and indeed they do.

The $X_{\mu}$, $X_{\mu \nu }$ can also be realized as 
differential operators on the fuzzy sphere $ \can ( S^{2k}) $. 
This is to be expected since we know that there is a
dual description in terms of  $U(n)$ gauge theory  on $S^4$ 
with instantons. 
In the  original matrix model, $Z_{\mu}$ can be thought of
as the matrix coordinates of D0-branes.

\subsec{ Derivations on $\can ( S^{2k} ) $ } 

 We showed above that the action by commutators of 
 $ X_{\mu } $ and $ X_{ \mu \nu } $, denoted 
 as $ ad ( X_{\mu} ) $ and $ ad( X_{\mu \nu } )$, 
  can be realized as 
 derivatives on the finite $n$ coset $\chan ( S^{2k} )$.   
 We show here that $ ad (X_{\mu \nu })$ also 
 acts as derivations on the non-associative 
 algebra $\can ( S^{2k} ) $, which is the finite 
 $n$ algebra of spherical harmonics. The product on this 
 space is defined by first multiplying as matrices and then 
 projecting onto symmetric 
 representations. 
 Indeed letting $A$ and $B$ be matrices in $ \can ( S^{2k} ) $,
 the product  $ A \circ B $ is defined by 
\eqn\dfpdca{ A \circ B = P ( A B ), } 
where $P$ acting on any matrix transforming in anything 
 other than a symmetric 
 representation is zero, 
and leaves invariant any matrix trasmforming in the symmetric
 traceless
 representation \fzalg. From its definition it is clear that 
 $P$ commutes with the action of $ SO(2k+1)$ on matrices. 
Let us denote $ ad ( X_{\mu \nu } ) = L_{ \mu \nu } $.
We can write  
\eqn\pfdr{\eqalign{
&L_{\mu \nu } ( A \circ B ) =    L_{\mu \nu }  ( P( AB ) ) \cr 
& =  P (  L_{\mu \nu }  ( AB ) )  
 = P (     L_{\mu \nu } ( A)   B + A    L_{\mu \nu } ( B ) )  \cr
& = P (  L_{\mu \nu } (  A ) B )   +  P ( A L_{\mu \nu } ( B ) )\cr 
& =  L_{\mu \nu } (  A ) \circ B + A \circ L_{\mu \nu } ( B ). \cr }}
This shows that $L_{\mu \nu } $ acts as a derivation 
on $ \can ( S^{2k} )$. This is a finite $n$ generalization of the 
action of $SO(2k+1)$ on functions on the sphere $S^{2k}$.

To obtain an action of the full set of generators 
entering the algebra \XXb, \XXc\ we can also work 
with just the sphere, without invoking the coset, but 
we need to equip the sphere with an embedding in ${\bf R}^5$ 
and consider vector fields on the normal bundle. 
We will describe the classical ( large $n$ )  version of this 
construction here. 

In terms of the Cartesian coordinates $x_{\mu}$
($\mu=1,\cdots, 2k+1$) in ${\bf R}^{2k+1}$,
functions restricted to a unit sphere are
functions of $z_{\mu}\equiv x_{\mu}/|x|$.
The naive derivative ${\partial\over\partial z_{\mu}}$
is not well defined because it does not respect
the constraint $z_{\mu}z_{\mu}=1$.

Define vector fields on the normal bundle by 
$$ \partial_{\mu}=|x|{\partial \over \partial x_{\mu}}. $$
We have
\eqn\Dz{ [\partial_{\mu}, z_{\nu}]=\delta_{\mu\nu}-z_{\mu}z_{\nu}
=P_{\mu\nu}, }
where $P_{\mu\nu}$ is the projection operator
which projects a vector in ${\bf R}^{2k+1}$ at $z$
to the tengent plane on $S^{2k}$.
One can check that the constraint $z_{\mu}z_{\mu}= 1$
is preserved under differentiation by $\partial_{\mu}$.
It turns out that
\eqn\DD{
[\partial_{\mu}, \partial_{\nu}]=
x_{\mu}{\partial\over\partial x_{\nu}}
-x_{\nu}{\partial\over\partial x_{\mu}}=L_{\mu\nu}, }
which is the generator of $SO(2k+1)$ rotations.
We see that $(\partial_{\mu}, L_{\mu\nu})$
satisfy the same relations \XXb\ and \XXc\
as $(X_{\mu}, X_{\mu\nu})$
(up to an overall factor of $2$).
 We will leave a finite generalization of this
 construction to the future. 



\newsec{ The fuzzy 4-sphere case } 
    
  Rather than analysing the complete set of equations and 
  looking for explicit solutions, a more efficient method is 
   to use the known matrices, and obtain classical quantities,  
   by taking expectation values between states. One can check
  afterwards 
 that the algebraic equations are indeed satisfied, as they should. 

 Take the state $ | s \> \equiv  |0\> \otimes |0\> \cdots |0\> $
 in $ Sym (V^{\otimes n} ) $ which 
 is the $n$-fold tensor product of the fermion Fock space vacuum.
 Since $ \Gamma_5 |0 \> = -|0\> $, we have $X_5 |s\> = -n |s\> $. 
 The normalized operator $Z_5$ has an expectation value
$ \< s| Z_5 |s\> = -1 $.
We also find that $Z_{12} = 1, Z_{34} = 1 $. 
So one solution to the system of algebraic equations 
that will be given in sec.4.4 is
\eqn\ones{\eqalign{&  Z_{5} = -1, \cr 
                   &  Z_{12 } = 1, \cr 
                   &  Z_{34} = 1, \cr }}
with all other $Z_{\mu} $ and $ Z_{\mu \nu } $ 
being zero. 
In the above we have abbreviated $\< s|Z|s \> $ as $Z$.

 Now we look at the subgroup of $SO(5) $ which 
 leaves this solution fixed. Requiring that the configuration  
 $X_{5} = - 1$,  $ X_i = 0 $ for $i=1,2,3,4$, 
 be left invariant leads to an $SO(4) $
 subgroup that acts non-trivially on the index $i$.   
 Consider linear combinations
$ \lambda = \sum_{ij} \lambda_{ij} L_{ij}$, 
 where the lambda matrix is displayed below : 
\eqn\matl{ 
\lambda =   \pmatrix{ 0  &   \lambda_{13}  & \lambda_{12} & \lambda_{14} \cr  
            -\lambda_{13} & 0    & \lambda_{14} & \lambda_{34}  \cr 
            -\lambda_{12} & -\lambda_{14} & 0 & \lambda_{13} \cr 
            -\lambda_{14} & -\lambda_{34} & -\lambda_{13} & 0 \cr }.
}
We have chosen for convenience to label the 
 rows and columns in the order $ (1324)$. 
These matrices $ \sum_{ij}  \lambda_{ij} L_{ij} $ take the form 
\eqn\utwo{ 
  A + B \pmatrix { 0 & 1 \cr 
                  -1 & 0 \cr }  = \pmatrix { A & B \cr 
                                   -B & A \cr }.  } 
Here $A$ is anti-symmetric and $B$ is symmetric. These 
$4 \times 4$ matrices form a $U(2)$ subgroup of  the 
$SO(4)$ subgroup of $SO(5)$ which leaves $X_5$ invariant.

 The generators  $L_{ij} $, which are antisymmetric matrices 
of the $SO(4)$ Lie algebra, 
 act 
 on $Z_{kl} $ as follows : 
\eqn\actL{ 
  L_{ij}(  Z_{kl}  )  = \delta_{jk} Z_{il} + \delta_{il} Z_{jk} -
 \delta_{ik} Z_{jl} - \delta_{jl} Z_{ik}.  } 
Consider a linear combination
$ \zeta = \sum_{i,j} \zeta_{ij} Z_{ij} $.
Let a linear combination  $\sum_{ij}  \lambda_{ij} L_{ij}$ 
 act on $ \zeta $. The result of this action is a sum $ \delta \zeta =
 \sum_{ij} V_{ij}
 Z_{ij} $. The coefficients $V$ depend bilinearly on $ \lambda$ 
 and $\zeta$. Considering $V_{ij} $ as 
the entries of a $ 4 \times  4 $ matrix $V$, \actL\ is equivalent 
to $ V = [ \lambda , \zeta ] $. Decomposing $ \zeta $ into $ 2 \times 2 $ 
blocks as 
\eqn\asmz{  \zeta = \pmatrix{ X & Y \cr 
                          -Y & \tilde X   },   } 
we have  $ X^{T} = -X $, $  {\tilde X}^{T} = - \tilde{X}
$ while  $Y$ is arbitrary. 
 For $ \lambda $ in the 
$U(2)$ subalgebra \utwo\   and using \asmz\ ,
\eqn\comtt{ V  = \pmatrix{ ([A,X] - [B,Y])  & (  [A,Y] + B
\tilde X - X  B ) \cr 
-BX  + {\tilde X } B - [A,Y ] & -[B,Y] + [A,\tilde X  ] \cr }. } 

Now we want to evaluate the variation $V$ 
on the solution we started with.
The solution can be described by a matrix $ \zeta^{(s)} $ which
is of the form 
\eqn\zijs{  \zeta^{(s)} =  \pmatrix{ 0 & 1  \cr 
                                1  & 0 \cr }, } 
where each entry is a $ 2\times 2 $ block, 
 and $0$ is the zero matrix, and $1$ is the identity 
 matrix.
 Given the form of the 
solution  \zijs, this is equal to the trace of the off-diagonal 
 blocks of $V$, i.e $ V_{12} + V_{34} $. 
Now $tr ( [A,Y] ) =0 $. Further,  
$B \tilde X $ and $BX $ are each antisymmetric, 
 given that $ B$ is symmetric and $ X, \tilde X$ are 
antisymmetric. Therefore, the off-diagonal block  of the variation 
$V$, when $ \lambda $ is in the $ u(2)$ subalgebra, has zero trace. 
This means that the solution is invariant under the $U(2)$ 
subgroup. 

We have therefore shown that the 
stabilizer group of the solution is 
$U(2)$. The action of $SO(5)$ then generates
a space of solutions which is $SO(5)/U(2)$. 
 The space related to the  large $N$ limit of the 
 matrix algebra is then  $ SO(5)/U(2)$. This is a bundle 
 over $S^4$ with fibre $ SO(4)/U(2)$, a symmetric space. 
 Using the isomorphism $ SO(4) = ( SU(2) \times SU(2) )/Z_2$ 
 it can be shown that the fibre is actually $ SU(2) / U(1)$, 
 i.e the two-sphere. 
 So we have in this case 
\eqn\fib{\eqalign{ 
     &  SO(5)/U(2) \longleftarrow SO(4)/U(2) \equiv SU(2)/U(1) \equiv S^2 \cr 
     & ~~ \downarrow ~~ \cr 
     & S^4 \equiv SO(5)/SO(4) \cr 
}}    

 We can also work directly with \actL\ and 
  show that the $U(2)$ subgroup 
described in \matl\ and \utwo\ leaves \ones\  invariant. 
The generators of $U(2)$ are
$L_{12}, L_{34}, L_{13}+L_{24}$, and $L_{14}-L_{23}$.
The generators $L_{13}-L_{24}$, $L_{12} + L_{34}$ and   $ L_{14}+L_{23}$
generate the $S^2$ bundle at $|s\> $ which we will explain 
in more detail in the next subsection. 

Another viewpoint toward this space is
to view $X_{\mu}$ and $X_{\mu\nu}$
as generators of the $SO(5,1)$ Lie algebra.
Similar arguments to the above lead to
the conclusion that the space of the matrix algebra
is the quotient space $SO(5,1)/U(2,1)\simeq SO(5)/U(2)$.

\subsec{ Bundle structure at finite $n$  }

In the previous subsection we evaluated the
operators $Z_{\mu}$ and $Z_{\mu \nu}$ for a state
$|s\> $ in $ Sym ( V^{\otimes n } ) $. 
By analysing the space of states which leaves 
the expectation values of $Z_{\mu}$ invariant, 
we will identify the space of states 
corresponding to the fibre of the 
non-commutative $SO(5)/U(2)$ coset over 
a fixed point of the sphere.

We first  define the base space of the bundle
in the following way.
As pointed out in \fzalg,
only totally symmetrized polynomials of $Z_{\mu}$
should be viewed as the quantum version
of functions on $S^4$.
Let us use the term {\it non-associative $S^4$ }
for the underlying space of
this (non-associative) algebra of functions.
This is our base space. The non-associative 
algebra of functions on the $S^4$ will be referred to 
as $ \can (S^4 )$. 

We define two states in the Hilbert space
$Sym(V^{\otimes n})$ of the matrix algebra
to be on the same fiber
if all functions in $\can ( S^4)$
have the same expectation values for both states.
According to the definition of Connes \connes,
this implies that the distance
\foot{ The distance between two states $\psi$, $\psi'$
on a space with algebra of functions ${\cal A}$ is
defined to be
\eqn\distance{ dist(\psi,\psi')=sup\{
|\< \psi|f|\psi\> -\< \psi'|f|\psi'\> | :
|Df|\leq 1, \; f\in {\cal A} \}, }
where $D$ is the so-called Dirac operator
which defines the differential calculus.
Although we did not specify the Dirac operator
for $\can(S^4 )$,
it should be clear that our comment here is valid
for any choice of $D$. }
between the projection of the two states on $\can(S^4)$
is zero.

The state $|s\> $, for example,
is on the same fiber with all (normalized) states 
in the space $ W \equiv  Sym ( V_{-}^{\otimes n }) $ where the 
space $V_{-} $ is spanned by $ |0 \> $ and $ a_1^{\dagger}
a_2^{\dagger} |0 \> $. 
One can check that $\< s'|Z_{\mu}|s'\> $
is the same for all $|s'\> \in Sym ( V_{-}^{\otimes n }) $. 
For $f$ in $\can (S^4)$,
$\< s'|f(Z)|s'\> $ is nonvanishing only if
it is a product of an even number of $Z_i$'s ($i=1,\cdots,4$).
In fact, the nonzero contributions are from
terms with even numbers of $\Gamma_i$'s
in each tensorial factor.
However, a total symmetrization will render
such products of $\Gamma_i$'s to be proportional
to the unity.
Since $\< v|\Gamma_5|v\> =-1$ for all normalized $|v\> $
in $V_-$.
We see that all functions on $\can (S^4)$
have the same expectation values for all $|s'\> $.

The states on the fiber $W$
form a representation for the operators
$M_1=Z_{13}-Z_{24}, M_2=Z_{23}+Z_{14}, M_3=Z_{12}+Z_{34}$,
which satisfy the algebra of a fuzzy $S^2$
\eqn\fiber{ [M_i, M_j]=-2i\epsilon_{ijk}M_k. }
The rest of the $Z_{\mu\nu}$'s have
vanishing expectation values on the fiber.
We can thus view the fiber as a fuzzy 2-sphere
with Cartesian coordinates $M_i$.
Since the matrix algebra is covariant under
$SO(5)$ transformations,
the same can be said about every point of 
the base space $ \can(S^4 )$. 
We conclude that the fuzzy 4-sphere
is in fact a fuzzy $S^2$ fiber bundle
over the base space of $ \can(S^4 )$. 

The points on the base sphere can be identified 
with the quotient space $ Sym ( V^{ \otimes n } ) / W $. 
The projection map from $Sym ( V^{ \otimes n } )$ 
to the quotient space is the finite $n$ analog of the 
projection map from the coset space $ SO(5)/U(2)$ to 
the sphere $S^4$.

Functions of $M_i$ are sufficient to distinguish
all states on the fiber.
In fact, $W$ is
a $(n+1)$ dimensional irreducible representation
for the $SU(2)$ Lie algebra generated by $M_i$.
The set of all functions of $M_i$ is
the same as the set of all $(n+1)\times(n+1)$ matrices.
Hermitian functions of $M_i$ generate
the Lie algebra of $U(n+1)$. This leads to a direct 
connection between the space of horizontal one-forms   ( with indices 
parallel to the base ) 
on the $SO(5)/U(2)$ coset with connections
of a $U(n+1)$ bundle over the sphere. Further, scalar 
fields on the coset correspond to adjoint scalars 
which are sections of the $U(n+1)$ bundle. 

\subsec{ Relation to unitary bundle over $S^4$  } 

Interpreting $X_{\mu\nu}\propto Z_{\mu\nu}$ 
as a field strength, as is familiar in 
Matrix Theory, the fuzzy  $S^2$ fiber bundle can
also be understood as a $U(n+1)$ vector bundle  on spherical D4-branes,
whose worldvolume is the non-associative $ S^4  $.

This connection can be used, in the large $n$ limit, to compare the 
$X_{\mu \nu }$ matrix at the point 
$ Z_{\mu} = ( 0,0,0,0,1)$ with the 
field strengths of the homogeneous 
instanton. We have $ X_{12} = X_{34}  $ 
while other field strengths are zero.
We will use the fact that
\eqn\gam{\eqalign{   
& 1=\< 0|\Gamma_{1}\Gamma_{2}|0\> =\< 0| \Gamma_{3}\Gamma_{4}|0\> , \cr 
& -1=\< 0|a_2a_1 \Gamma_{1} \Gamma_{2} a_1^{\dagger}
 a_2^{\dagger} |0\> =\< 0|a_2a_1 \Gamma_{3} \Gamma_{4} a_1^{\dagger}
 a_2^{\dagger} |0\> ,  \cr  }} 
A basis in $Sym ( V_{-}^{\otimes n} )$ can be labelled by the 
number of $|0\> $'s appearing in the tensor product. 
In such a basis it is clear that $ X_{12} = X_{34}$ 
is proportional to
\eqn\Diag{ Diag  ( n, n-2, \cdots ,  -n+2  , -n). }
Compare this with 
the solution described in Appendix B of  \cmt\ where the $F_{12} =F_{34}$ 
are the only non-zero field strengths at the point $ \alpha_1 = {\pi
\over 2 }  $ with all other angles zero. They are proportional 
to the embedding of $2 \sigma_3$ in $U(n+1)$ using the standard $(n+1)$
dimensional representation of $SU(2)$, which is again 
the matrix \Diag\ .

A physical interpretation of this space is the following.
The $Z_{\mu}$'s can be viewed as matrix coodinates
of $N\simeq n^3/6$ D0-branes. We know that the solution 
given by  matrices \actX\ also contains spherical 4-branes. 
The number of these 4-branes at large $n$ can be computed using 
the definition of 4-brane charge in Matrix Theory \refs{ \grt\bss } 
as done in \clt. In the application considered by 
\cmt\ the zero-branes are replaced by $1$ branes and 
the 4-branes by $5$-branes. The number of 5-branes can  be computed, 
in a manner similar to the Matrix Theory application,  
by analyzing couplings in the D1-brane action.
Note that, in the previous paragraph, 
 the number of $D4$ ( or $D5$ ) was extracted 
directly from the fuzzy fibre bundle structure of the 
$ \chan ( S^{2k} ) $ without appealing to couplings on brane 
actions, a fact which should be useful in more general contexts 
in Matrix Theory and D-brane physics.   

In the dual description, on the worldvolume 
of $n$ coincident spherical D4-branes, we have a $U(n)$ gauge theory,  
$X_{\mu}$'s are interpreted as the $U(n)$ covariant derivatives
and $X_{\mu\nu}$ as field strengths. 
D0-branes are realized as instantons of the $U(n)$ gauge field.
In \ones\ we see that the field strength $X_{\mu\nu}$
is indeed a self-dual configuration at the north pole.
Since the matrix algebra is invariant under $SO(5)$,
it is clear that the matrices $X_{\mu\nu}$ correspond
to homogeneous instantons on $S^4$.
It was found in \wang \cmt\ that the maximal number
of homogeneous instantons for an $U(n)$ bundle on $S^4$
is $n(n^2-1)/6$.
In the large $n$ limit, 
this is precisely the instanton number for our configuration $X$, 
as first observed in \cmt.

Fluctuations around this configuration can be
described in terms of fields on the coset $SO(5)/U(2)$.
If a transverse matrix coordinate $\Phi$
is excited to have a dependence on $Z_{\mu\nu}$,
such as $\Phi\sim Z_{\mu\nu}$,
a D2-brane dipole with charge density
$tr(Z_{\mu}Z_{\nu}\Phi)$ is generated \cmt\ .
This is because each D4-brane has a different
D2-brane charge density. They add up to zero 
when the D4-branes are on top of each other.
But if different D4-branes have
different transverse fluctuations,
D2-brane dipoles will appear.

\subsec{ Traces and Integration } 

Integration on this non-commutative coset $SO(5)/U(2)$
can be defined as the trace over $Sym(V^{\otimes n})$.
The cyclicity of the trace automatically ensures
$SO(5)$ invariance of the integration because
$$ Tr(L_{\mu \nu }(f(Z)))={1\over 2}Tr([X_{\mu \nu },f(Z)])=0. $$
For a given function of $Z$,
one can decompose it into irreducible representations
of $SO(5)$ as \fzalg\
$$ f(Z)=\sum_{R,s} a_{R,s} Y_{R,s}(Z), $$
where $R$ stands for irreducible representations
and $s$ for the indices of states in $R$.
The $SO(5)$ invariance implies that
$$ \int d \Omega f(Z) = a_{0} = { 1 \over N } Tr ( f )  , $$
where $a_0$ is the coefficient for
the trivial representation, which is unique
(with multiplicity one). 
Hence, in the large $n$ limit,
this integration agrees
with the usual integration on $SO(5)/U(2)$
induced from the unique Haar measure on $SO(5)$.
The measure $ d \Omega $ is normalized such that 
the volume of the coset is $1$.

\subsec{ Further remarks on the coset $SO(5)/U(2)$ } 
             
Here we provide another way to obtain
the same result of the coset $SO(5)/U(2)\simeq S^4\times S^2$
by directly investigating the matrix algebra.
Some of the algebraic relations for the large $n$ matrix algebra
are 
\eqn\radius{ Z_{\mu}Z_{\mu}\simeq {\bf 1}, }
\eqn\Zmunu{ Z_{\mu\nu}Z_{\mu\nu}\simeq 4\times{\bf 1}, }
\eqn\consa{ \epsilon_{\mu\nu\lambda\rho\kappa}
Z_{\mu\nu}Z_{\lambda\rho}\simeq -8Z_{\kappa}, }
\eqn\consb{  Z_{\mu\nu} Z_{\nu\lambda}
\simeq -2\delta_{\mu\lambda}+ Z_{\mu}Z_{\lambda},  }
\eqn\consc{ Z_{\mu} Z_{\mu \lambda } \simeq 0. } 
One can also show that
\eqn\new{
\epsilon_{\mu\nu\lambda\rho\kappa}Z_{\mu}Z_{\nu\lambda}
\simeq 2Z_{\rho\kappa}, }
which says that $Z_{\mu\nu}$ is a self-adjoint
tensor field on the sphere.

Relation \radius\ means that $Z_{\mu}$
can be viewed as Cartesian coordinates
of a unit fuzzy sphere.
The identity \Zmunu\ implies that the magnitude
$| Z_{\mu\nu} |$ is of order $1$.
Incidentally, since
\eqn\ZZZ{ [Z_{\mu}, Z_{\nu}]= {-2i\over n}Z_{\mu\nu}, }
we can estimate the uncertainty in $Z_{\mu}$ as
\eqn\UR{ \Delta Z_{\mu} \Delta Z_{\nu} \sim {\cal O}(1/n).}
This means that ``points'' on the 
the unit fuzzy sphere with Cartesian coordinates $Z_{\mu}$
have spread which scales like $1/n$.

In the large $n$ limit,  the commutator of 
$Z_{\mu}$ \ZZZ\ vanishes, showing that  
$Z_{\mu\nu}$ has to be included as generators
of the matrix algebra. 
All functions of $Z_{\mu}$'s can be
obtained from totally symmetrized products
of $Z_{\mu}$'s and $Z_{\mu\nu}$'s.
Yet the $Z_{\mu\nu}$'s are not completely independent,
but are constrained by \Zmunu, \consa\ and \new\ .
Up to terms of order ${\cal O}(1/n^2)$,
we can interpret these relations as
constraints on a classical space with commutative coordinates
$Z_{\mu}, Z_{\mu\nu}$.
At a given point on $S^4$,
say, $(Z_1,\cdots,Z_5)=(0,0,0,0,1)$,
the space of solutions to these constraints
is the set of anti-self dual tensors $Z_{ij}$
for $i,j=1,2,3,4$. This space is a 2-sphere.
The whole space is thus locally $S^4\times S^2$.

\newsec{ The fuzzy 6-sphere case } 

\subsec{ Solution to Equations and Stabilizer groups } 

Analogous to \ones, the expectation values of $|s\> $ are
\eqn\sxcls{
\eqalign{  & Z_7 = -1 , \hbox{ and} ~ 
Z_i = 0 ~ \hbox{ for } i = 1 \cdots 6, \cr 
            & Z_{12} = Z_{34} = Z_{56} = 
{ 1 \over \sqrt 6}  ~ \hbox{ and} ~  Z_{\mu \nu }  = 0 ~ 
 \hbox{ otherwise},  \cr 
 & Z_{127} = Z_{347} = Z_{567 } = { 1 \over \sqrt{ 18} }  
~ \hbox{ and}~  
 Z_{\mu \nu \rho} = 0 \hbox
 { otherwise }. \cr }}
 We have here defined $Z$'s so that they satisfy 
 the normalization condition 
\eqn\norms{
\sum_{\mu}Z^2_{\mu}=1,\quad
\sum_{\mu\neq\nu}Z^2_{\mu\nu}=1,\quad
\sum_{\mu  \ne \nu  \ne \rho } Z_{\mu \nu \rho }^2 = 1. } 
The $Z_{ \mu \nu \rho}$'s are antisymmetric in the three indices. 
Eq.\sxcls\ is a classical solution to 
\norms\ and self-adjoint relations analogous to \consa\ and \new.

We prove that the stabilizer group of this solution
is $U(3)$.  We define $\zeta = \sum_{ij} \zeta_{ij} Z_{ij} $. 
On the classical solution this takes the value 
\eqn\valmat{  \zeta^{(s)} = { 1 \over \sqrt 6 }
\pmatrix{ 0 & 1 \cr 
          1 & 0 \cr },
} 
 where we have chosen to label the rows and columns of the 
 matrix as $ ( 135 246 )$. 
Similarly we can define a $\zeta_7 = \sum_{ij} \zeta_{ij} Z_{ij7}$. 
And we have after evaluating on the solution 
\eqn\zesev{   \zeta_7^{(s)}  =  { 1 \over \sqrt {18}}
  \pmatrix{ 0 & 1 \cr 
            1 & 0 \cr }.  } 

The variations of $\zeta$ and $\zeta_7$ are both 
obtained by the action of commutators as in the case of fuzzy $S^4$. 
 Therefore an argument
similar to the one above shows that $U(3)$ Lie algebra matrices 
of the form 
$$ \pmatrix{ A & B \cr 
             -B & A \cr } $$
with $A$ antisymmetric and $B$ symmetric leave the solution
invariant. As before the final step will rely on the tracelessness
of the off-diagonal blocks of matrices of the form \comtt, 
where the blocks are now $ 3 \times 3$ matrices.

So the stabilizer group is $U(3)$. By considering the action of 
the $SO(7)$ symmetry of the equations satisfied by the variables 
$ Z_{\mu}$, $ Z_{\mu \nu }$, and $ Z_{ \mu \nu \lambda} $, 
 we find that a space $ SO(7)/U(3)$ is generated. 
 This is a bundle over the sphere $ S^6$. The fibre is 
 $SO(6)/U(3)$. 
\eqn\fibss{\eqalign{ 
     &  SO(7)/U(3) \longleftarrow SO(6)/U(3) \cr 
     & ~~ \downarrow ~~ \cr 
     & S^6 \equiv SO(7)/SO(6) \cr 
}}    

\subsec{ Bundle structure at finite $n$ } 

The fibre over the sphere $S^6$ is, in the 
classical limit, the symmetric space $SO(6)/U(3)$. 
At finite $n$ we have a fuzzy version of 
this space. 

As in the discussion of $ \chan( S^4)$, 
the vector subspace 
$ Sym ( V_-^{\otimes n } ) $ can be identified 
with the points on the fibre. This space is 
spanned by $ |0 \> $, and four  states of the form 
$ a_i^{\dagger} a_j^{\dagger} |0 \> $. The symmetric 
product is an irreducible representation of $SO(6)$ 
with weights $ \vec \lambda = ( { n \over 2 } , { n \over 2 } , {n
\over 2 }  )$. This space has dimension ( see for exampe \fulhar\ ) 
\eqn\dimspr{ D_6 =  { 1 \over 6 } ( n+1) ( n+2)  (n+3). } 
The matrix algebra over this space 
is generated by the operators $Z_{\mu\nu}$. 
Multiplying these operators leads to 
operators of the form 
\eqn\opfrm{ \sum_{s_1,s_2,\cdots s_r } \rho_{s_1}  ( \Ga \Ga ) 
\rho_{s_2} ( \Ga \Ga ) \cdots \rho_{s_r} (  \Ga \Ga ). } 
This is somewhat schematic since we have 
not written out the indices on the $\Ga$ matrices. 
To be more precise we would write out the indices 
and contract with a traceless tensor which has the 
symmetries of the Young diagram with two rows of length 
$r$. Every pair of $\Ga$ in the same $ \rho $ factor 
is antisymmetrized. 
Such a representation has dimension 
\eqn\dimtwr{ D(r) = { 1 \over 12 } ( 2 r +3) (r+2)^2 (r+1)^2. } 
By adding up these dimensions in the range
$0 \le r \le n $ we get 
\eqn\sm{ \sum_{r=0}^{n} D(r) = D_6^2. }

\subsec{ Relation to unitary bundles over the sphere } 

All these matrices are hermitian, so they 
form a basis for the Lie algebra of the
unitary group with rank  $ D_6 = { 1 \over 6 } ( n+1) ( n+2) 
 (n+3)$.
This indicates that the number of spherical 
6-branes when we use $ \Phi_{\mu} \sim L Z_{\mu }$ 
is ${ 1 \over 6 } ( n+1) ( n+2) (n+3)$,
giving a purely non-commutative geometric 
derivation of the 6-brane charge. 
As before, the $Z_{\mu\nu}$ can be interpreted
as field strengths for unitary bundles.

\newsec{ The fuzzy $8$-sphere case }

\subsec{ Solutions to equations and stabilizer group } 
 The large $N$ limit of the matrix algebra is generated 
 by commuting variables, $ Z_{\mu} $, $ Z_{\mu \nu } $, $ Z_{\mu \nu
 \lambda}$ and $ Z_{\mu \nu \lambda \rho} $. The indices run from 
 $1$ to $9$ and any variable with more than one  index  is
 antisymmetric under exchange of any pair of indices. 
 
 By taking the expectation values of the appropriate matrices in 
 a state one finds a particular solution : 
\eqn\onese{\eqalign{ 
 & Z_{9} = 1, \cr
 & ~~~~~ \hbox{ and } ~Z_{i} = 0 ~ \hbox{ for all other}~  i, \cr 
 & Z_{12} = Z_{34} = Z_{56} = Z_{78} = 
{ 1 \over \sqrt{8}}, ~ \cr 
& ~~~~~ \hbox{and} ~Z_{\mu \nu} = 0 ~  
\hbox{ otherwise }, \cr 
& Z_{129} = Z_{349} = Z_{569} = Z_{789} =
{ 1 \over { \sqrt{ 24} } }, \cr 
& ~~~~~ \hbox{ and }~ Z_{\mu \nu \rho } = 0~ \hbox{ otherwise }, \cr 
 & Z_{1234} = Z_{1256} = Z_{1278} = Z_{3456} = Z_{3478} = Z_{5678} =
 {-1\over 12} ~~ ,\cr 
& ~~~~~ \hbox{ and } Z_{\mu \nu \rho \lambda} = 0 ~ 
\hbox{otherwise}.\cr }}
 By the statement that the variables with more than one index 
 are zero otherwise, we mean that all components not related to 
 the ones shown by the permutation symmetries are zero. 
 For example $ Z_{219} = - Z_{912} = - { 1 \over { \sqrt{ 24} } }$, 
 but $ Z_{139} = 0 $.

The normalizations are chosen such that 
\eqn\norms{\eqalign{&  \sum_{\mu} Z_{\mu }^2 = 1, \cr 
                    &   \sum_{ [\mu ,\nu] } Z_{ \mu \nu }^2 = 1, \cr 
                    &  \sum_{ [\mu, \nu, \lambda ] } Z_{\mu \nu
\lambda }^2 = 1, \cr 
 & \sum_{ [\mu, \nu, \lambda,\rho ] }
Z_{[ \mu \nu \lambda \rho] }^2 = 1. \cr }} 
 The sums  $ [\mu , \nu ]$   etc.  indicate that the indices 
 run over $1$ to $9$ while respecting the condition 
 that they are never equal. These variables are not 
 all independent and obey contraints like the one following from \SD, 
 and others that can be derived from the definitions in section 3.2
 in the large $n$ limit.  
  
  Now we consider the subgroup of $SO(9)$ which leaves the 
  solution invariant, i.e the stabilier group.
 Requiring that the $Z_{\mu}$ values stay
  invariant clearly forces the stabilizer group to be a subgroup 
 of $SO(8)$. Now requiring the $Z_{\mu \nu}$ and $Z_{\mu \nu \lambda}$ 
 to be invariant picks out a $U(4)$ subgroup of the $SO(8)$. The proof 
 proceeds as for the case of the fuzzy $S^6$. 
 This can be described by taking $ 8 \times 8 $ matrices of the form 
 $$ \pmatrix{ A & B \cr 
              -B & A \cr },$$
 where $B$ is symmetric and $A$ is antisymmetric, and 
 the rows and columns are labelled in the order $(13572468)$. 
  We still need to prove that the configuration of 
 $Z_{\mu \nu \lambda \rho}$ is invariant and this can be done directly
 using the following $SO(9)$ transformations of 
the $Z_{\mu \nu \lambda \rho}$
\eqn\zftns{\eqalign{ 
& L_{\mu_1 \mu_{2} } ( Z_{\nu_1 \nu_2 \nu_3 \nu_4} ) 
 = ( \delta_{\mu_2 \nu_1} Z_{\mu_1 \nu_2 \nu_3 \nu_4} 
    - \delta_{\mu_2 \nu_2 }  Z_{\mu_1 \nu_1 \nu_3 \nu_4} 
     + \delta_{\mu_2 \nu_3 }  Z_{\mu_1 \nu_1 \nu_2 \nu_4} 
     - \delta_{\mu_2 \nu_4 }  Z_{\mu_1 \nu_1 \nu_2 \nu_3} ) \cr 
&     - ( \mu_1 \leftrightarrow \mu_2 ). \cr }}

The explicit proof is tedious, we just show some example steps. 
Take for example the combination $L_{13}  + L_{24}$
which is part of the $U(3)$ subgroup. We have
\eqn\actlc{\eqalign{ 
           & L_{13} ( Z_{3256} ) = Z_{1256}, \cr   
           & L_{24} ( Z_{3256} ) = - Z_{3456}. \cr }} 
Since $Z_{1256} = Z_{3456}$ on the solution, this means 
 that the variation by $ L_{13} + L_{24} $ on $ Z_{3256}$ 
 is zero. Take another example, the variation of $Z_{1356} $. 
 Now $L_{13} ( Z_{1356}) = L_{24} ( Z_{1356}) $ are both 
 seen to be zero using \zftns\ even before evluating on the solution. 
 Another example is the variation of $ Z_{3456} $. The action 
 of $L_{13}$ on this is $ Z_{1456}$ which is zero on the solution, 
 and the variation of $L_{24}$ gives $ Z_{3156}$ which is also zero. 
 Similarly we can check for each generator of the $U(4)$ that 
 the variation of any of the $Z_{\mu \nu \lambda \rho }$ coordinates 
 is zero. The steps are obvious and we will not spell out the complete 
 proof.

 Having established that the stabilizer group 
 is $U(4)$, it follows that action of the $SO(9)$ 
 symmetry group generates a coset $SO(9)/U(4)$. 
 This is a bundle over the $8$-sphere. 
\eqn\fibse{\eqalign{ 
     &  SO(9)/U(4) \longleftarrow SO(8)/U(4) \cr 
     & ~~ \downarrow ~~ \cr 
     & S^8 \equiv SO(9)/SO(8) \cr 
}}

\subsec{ Bundle structure 
of the coset at finite $n$ } 

Viewed as a bundle over $S^8$, the coset 
$SO(9)/U(4)$ has a fibre $SO(8)/U(4)$. 
At finite $n$ we have, as a space of states 
which has the same expectation values 
for $X_1, \cdots X_9$, the 
space $ Sym( V_-^{\otimes  ~n } )$, where $V_-$ is 
spanned by states of the form $ |0 \> $, 
$ a_{i}^{\dagger}  a_{j}^{\dagger} |0\> $ 
and $ a_{1}^{\dagger} a_2^{\dagger} a_{3}^{\dagger} 
   a_4^{\dagger} |0 \> $. This symmetric 
 tensor product space is an irreducible representation 
of $SO(8)$ with weight $ \lambda = ( { n \over 2 } ,  { n \over 2 }, 
    { n \over 2 },  { n \over 2 } )$ and dimension \fulhar\ 
\eqn\dimsp{ 
D_8 =  { 1\over 360 } ( n+5) (n+4) (n+3)^2 (n+2 ) (n+1). } 

The operators $Z_{\mu\nu} $ and $ Z_{\mu\nu\lambda\rho} $ 
generate the matrices acting on this 
space. We get operators of the form 
\eqn\frmopii{ 
\sum_{ \vec s , \vec t } \rho_{s_1} ( \Ga \Ga \Ga \Ga ) \cdots 
\rho_{s_{p_1} } ( \Ga \Ga \Ga \Ga ) ~~ 
  \rho_{t_1} ( \Ga \Ga ) \cdots 
\rho_{t_{p_2} } ( \Ga \Ga ). } 

After endowing operators of this form with $SO(8)$ 
indices and contracting with traceless tensors of the appropriate 
symmetry, these correspond to irreducible representations 
associated with Young diagrams having  row lengths 
$ (p_1 +p_2, p_1 +p_2, p_1, p_1 )$.
This representation has dimension 
\eqn\redp{\eqalign{ 
&  D(p_1,p_2 ) = { 1 \over 4320 } ( 2p_1 +2p_2 +5) ( p_2 +2)^2 
  (p_2 +2p_1 +4 ) ( p_2+3)\times \cr 
&  ~~~~~ \times ( p_2+3 +2p_1)^2 (  p_2+1 ) (p_2 + 2p_1
+ 2  ) ( 2p_1 +1).  } }  
Adding up these dimensions, using Maple,  we find 
\eqn\add{ \sum_{p_1=0}^{n} \sum_{p_2=0}^{n-p_1} D(p_1,p_2 ) 
 = D_8^2, }   
where $D_8$ is given in \dimsp. 

These formulae allow us to 
read off the representations of $SO(8)$  
and  their multiplicities ( all $1$ ) appearing as 
 harmonics on the coset space $SO(8)/U(4)$.

\subsec{ Relation to unitary bundles }

The generating operators $Z_{\mu\nu}$ 
and $Z_{\mu\nu\lambda\rho}$ are hermitian and their products
which generate the full matrix algebra of 
dimension $D_8^2$ are all hermitian. So they
 can be thought as Lie algebra elements 
for $U(D)$. This shows that a fuzzy 8-sphere 
construction based on these matrices 
in any matrix brane action involves 
$D_8$ spherical 8-branes. 
As we discussed in the section on the 
fuzzy 4-sphere, we can read off 
the  $U(D) $ field strengths  at the N-pole 
of the sphere.

\newsec{ Action for fluctuations and field theory 
on  $SO(2k+1)/U(k)$ } 

We will outline here some features 
of the action describing fluctuations 
around classical solutions of matrix brane world-volume 
theories, such as the solutions in \cmt, \ho\ or \clt. 
The details will depend on the particular application. 
We will be somewhat schematic, outlining 
generic features, in particular the appearance of 
the $SO(2k+1)/U(k)$ coset as the base space of a field theory
having $U(1)$ gauge fields and an action compatible 
with the geometrical $SO(2k+1)$ symmetry. 
For concreteness we will discuss the case of 
the four-sphere with $SO(5)$ symmetry.

 Consider either the  application of \cmt\  to 
 brane intersections or the application in \clt. 
 Take any scalar transverse to the $5$ scalars that 
 go into the fuzzy sphere. Call it $\Phi$. 
 We expand it as $ \Phi = \sum_{R,s} a_{R,s } Y_{R,s} ( Z ) $. 
  The $Y_{R,s} $'s are operators associated 
   with the state $s$  in the  representation $R$ 
  of $SO(5)$. They are normalized as
\eqn\norm{ 
\int d \Omega ~~  Y_{R,s} ( Z) Y_{R,s} ( Z^{\dagger} )  = 1 .} 
To ensure correct normalization, operators 
 of the form 
\eqn\frmg{\eqalign{ 
 \sum_{\mu^i_j} A[ \mu^i_j]
\sum_{ \vec s } &\rho_{s_1} (\Ga^{\mu^1_1}
\Ga^{\mu^2_1}) \rho_{s_2} (\Ga^{\mu^1_2} \Ga^{\mu^2_2})   \cdots 
 \rho_{s_{r_2}}  ( \Ga^{\mu^1_{r_2} } \Ga^{\mu^2_{r_2}} ) \times \cr 
&\times \rho_{s_{r_2 +1 }} ( \Ga^{\mu^1_{r_2+1}} ) \cdots 
\rho_{s_{r_1 }} ( \Ga^{\mu^1_{r_1 } } ),  \cr }} 
associated with Young diagrams having a first 
row of length $r_1$ and a second row of length $r_2$,
are multiplied by a factor $ N (n, r_1,r_2)$, 
 which behaves in the large $n$ limit as $ n^{-r_1 }$.

 From the kinetic term for $\Phi$ we obtain    
\eqn\timed{  { 1 \over g_sl_s  }
\int dt~ Tr ~ ( \partial_t \Phi )^2 ~ 
  = { N  \over g_s l_s } 
 \int dt d\Omega  ~~( \partial_t \Phi )^2. } 
 On the right hand side
we have converted the trace into an integral, 
 and we have recognized $ \Phi$ as a field living 
 on the coset. 

 Other terms in the action give rise to 
 spatial derivatives acting on the field $ \Phi$. 
Indeed we have terms in the zero-brane action 
which are of the form
$ {1 \over g_s l_s^5} \int dt Tr [ \Phi_i , \Phi ]^2$. 
We expand $ \Phi_i =  L  Z_i + L^2 A_i$, where
$L$ is the size of the sphere described by the matrix 
coordinates $\Phi_i$ of the zero branes and  
$A_i = \sum_{R,s}  a_i^{R,s}  Y_{R,s}  ( Z ) $. 
We chose a factor $L^2$ in front of $A_i$ 
to ensure that it has the dimensions 
 of a gauge field. Since we identiftied 
 $X_i$ as derivatives section 3.3 we will get terms 
 of the form 
\eqn\spad{ 
{-  N \over g_s l_s^5 } \int dt d \Omega 
~~  \left( { L \over n }\left( Z_j { \partial \over  \partial Z_{ij} }+ 
 Z_{ij}{ \partial \over  \partial Z_{j} } \right) \Phi \right)^2.  }
We also have terms of the form 
\eqn\com{ {  L^4 \over g_sl_s^5 } \int dt ~ d \Omega
( [ A_{i}, \Phi ])^2. }
  This should be interpreted as a commutator 
 of star products. More detailed formulae for the 
 star product can be developed  along the lines of \iso\ 
 for example.

Similarly we have terms of the form 
 $  {1 \over g_s l_s^5}  \int dt ~ Tr ~ ( [ \Phi_i, \Phi_j ] ) ^2 $. 
These lead to expressions which include
 $\int dt d \Omega  ( D A )^2$, 
 $  \int dt d \Omega D^2$ and $ \int [ A,A]^2 $ terms, 
where the $D$'s are appropriate derivatives which can be read off from 
section 3.   
The term $\int dt d \Omega D^2 $ is just constant. 
The remaining terms are all terms we may expect from a
theory containing gauge fields and scalars living on $SO(5)/U(2)$ with 
an $SO(5)$ symmetry and a non-commutative $U(1)$ gauge symmetry. 

The non-commutative $U(1)$  gauge symmetry can be derived from the 
original symmetries of the Matrix Theory. We know that 
there is a unitary symmetry generated by hermitian 
 matrices $ \Lambda$. Since hermitian matrices give the 
spherical harmonics on the $SO(5)/U(2)$ coset, we 
 have symmetry variations of the form 
\eqn\syms{  \delta_{\Lambda}  \Phi_i = [ \Lambda, \Phi_i ], } 
which translate into 
\eqn\transsym{ \delta_{\Lambda } A_i = [ X_i, \Lambda ] + [A_i,
\lambda],  }  
where $ \Lambda $ is a function on the coset.

Note, however, that this $U(1)$ gauge theory
is different from ordinary gauge theories
in that the gauge potential $A_i$ is defined
only for directions along the non-associative sphere,
not for all directions on the coset.
This is because there are only $2k+1$ matrix coordinates
$X_{\mu}$ in the original matrix model,
which are Cartesian coordinates on a $2k$-sphere.
The $k(k-1)$ new dimensions corresponding to
fibres of the coset $SO(2k+1)/U(k)$
are generated by noncommutativity,
which is dictated by the equations of motion
for the matrix coordinates.
This is reminiscent of the recent proposal
of dynamically generated new dimensions \AH.

\newsec{ Summary and outlook }

 We identified some higher dimensional 
 geometries which are relevant to the 
 matrix construction of fuzzy spheres 
 of dimension greater than two. When 
 we look at fuzzy 4-spheres, the relevant 
 geometry is $6$-dimensional. For $6$-spheres
 it is $12$ dimensional, and for $8$-spheres 
 it is $20$ dimensional. 
  These geometries of the form $SO(2k+1)/U(k)$
 are bundles over spheres $S^{2k}$ with 
 fibre $ SO(2k)/U(k)$.

 We outlined an approach to understand this geometry 
 in terms of constructing 4-branes by locally 
 putting together D0-branes, in the case of the fuzzy 4-sphere.
 It is shown that the fuzzy coset $SO(5)/U(2)$ 
 can be viewed locally as the product of fuzzy 2-sphere
 and non-associative 4-sphere.
 We generalized these to the 6-sphere and 8-sphere.

 In developing the structure of the fuzzy $SO(2k+1)/U(k)$
 algebras as bundles over a non-associative sphere
 with a fuzzy $SO(2k)/U(k)$ fibre,
 we were lead to an interpretation of the coordinates 
 on the fuzzy fibre as field strengths on the base sphere 
 for a unitary bundle whose rank could be read off from 
 the geometry of the fuzzy fibre space. This gives a
 purely non-commutative geometric derivation of the brane 
 charges involved. It also allows us to construct directly 
 from the matrices the field strengths for the interpretation 
 in terms of field theory on a sphere. For the case of the 
 four sphere, this allowed us to reproduce the field strengths 
 of the homogeneous instantons first discussed in this 
 context in \cmt. It would be interesting to develop this
 further to reproduce not just the field strength
but also the connection. 
We find that the fuzzy spheres
provide very peculiar examples of noncommutative space
for which the noncommutativity of the non-Abelian gauge group
is mixed with that of the base space.

 Further study of the higher dimensional cases 
 would also be interesting. The $SO(2k+1)$ covariant 
 Matrix Theory construction should allow the 
 reconstruction of classical solutions for 
 non-Abelian theories on $S^{2k}$ in the large 
 $N$ limit.

 The fluctuations of the solutions have a 
description in terms of non-Abelian theory
of multiple branes on $S^{2k}$.
Interestingly, there is also 
a description in terms of an Abelian theory 
on the {\it higher} dimensional coset space 
$SO(2k+1)/U(k)$. 
We outlined some elements of this description, 
noting the $U(1)$ gauge symmetry and $SO(2k+1)$
 invariance. 
 The detailed geometry of this 
 Lagrangian remains to be understood. 
 More generally, it 
 will be interesting to further understand and interpret the 
 appearance of these higher dimensional 
 geometries in string theory.  

 While the fuzzy 4-sphere has been studied 
 in some detail \refs{ \clt, \cmt, \ps } in the 
 context of Matrix constructions, involving a
 low-dimensional brane growing into a higher dimensional brane, 
 in contexts where several physical points of view 
 are available, e.g the lower dimensional brane 
 world-volume, the spherical brane worldvolume as 
 well as supergravity, it will be very instructive 
 to conduct analogous studies of more general spheres. 
 Such generalizations should shed further light on the 
 physics of the higher coset geometries  $ SO(2k+1)/U(k)$.   
 Another potential class of applications of the various 
 non-commutative geometries we discussed 
 is  purely field theoretic  and attempts to find 
 symmetry preserving finite approximations 
 to base spaces of field theories. Some 
 recent work on different examples of 
 non-commutative  cosets appears in 
 \bdlmo\tv, and metric aspects of the non-commutative 
 geometry of spheres  are studied in \rief.

\bigskip
\noindent{\bf Acknowledgements:}
 We wish to thank for pleasant discussions Steve Corley, 
 Vladimir Fock, Zack Guralnik, Antal Jevicki,   
  David Lowe and Djorje Minic. We also thank P. Etingof and  I. Singer
 for email communications.  
 The research of S.R. was supported in part by DOE grant  
 DE-FG02/19ER40688-(Task A). 
The research of P.M.H. was supported in part by
the National Science Council, R.O.C.,
the Center for Theoretical Physics
at National Taiwan University,
the National Center for Theoretical Sciences,
and the CosPA project of the Ministry of Education,
Taiwan, R.O.C.

\newsec{ Appendix : Jordan algebras and fuzzy spheres } 

We describe some further algebraic properties
of $ \can  ( S^{2k} )$, $ k \ge 2 $. For concreteness we describe 
the case  $ \can  ( S^{4} )$, but the arguments are general. 

We first prove that the multiplication in $ \can( S^{4} )$
is commutative. 
Let us begin with some examples:
\eqn\exprod{\eqalign{  
&  X_{\mu} \circ X_{\nu  } = P ( \sum_{r} \rho_r( \Gamma_{\mu} ) \rho_s
( \Gamma_{\nu } ) ) \cr 
& =  P ( \sum_{r} \rho_r( \Gamma_{\mu} \Gamma_{\nu}  ) ) 
 + P ( \sum_{r \ne s }  \rho_r( \Gamma_{\mu} ) \rho_s( \Gamma_{\nu}  ) )
\cr 
& =  \sum_{r} \rho_r( \delta_{\mu \nu } ) + 
   \sum_{r \ne s }  \rho_r( \Gamma_{\mu} ) \rho_s( \Gamma_{\nu}  ) \cr 
& = n \delta_{\mu \nu } +   
 \sum_{r \ne s }  \rho_r( \Gamma_{\mu} ) \rho_s( \Gamma_{\nu}  ) \cr 
& = { ( n^2 + 4n ) \over 5 }   \delta_{\mu \nu } 
  + X_{( \mu \nu ) }. \cr}}
In the third  line, we have used the fact that the antisymmetric 
part of the first term belongs to a representation with 
row lengths $ \vec  r = ( 1,1 )$ which is projected out by $P$. 
This leaves only the symmetric part which is proportional 
to $ \delta_{\mu \nu } $. In the last line, we have introduced
a symmetric traceless tensor of rank $2$
\eqn\symdef{   X_{( \mu \nu ) } =
\sum_{r \ne s }  \rho_r( \Gamma_{\mu} ) \rho_s( \Gamma_{\nu}  )
- \delta_{\mu \nu } { n(n-1) \over 5 }. } 

It is clear that the product in \exprod\ is commutative. 
In general we have a product of the form 
\eqn\gfprod{\eqalign{ 
& X_{(\mu_1 \mu_2 \cdots \mu_k) } X_{ (\nu_1 \nu_2 \cdots \nu_l )} 
= a_1 X_{  ( \mu_1  \mu_2 \cdots \mu_k\nu_1 \nu_2 \cdots \nu_l ) } \cr 
&   + a_2 ( \delta_{\mu_1 \nu_1}  
 X_{  (  \mu_2 \cdots \mu_k  \nu_2 \cdots \nu_l ) } + 
 \delta_{\mu_2 \nu_2}
X_{  (  \mu_1 \mu_3 \cdots \mu_k  \nu_1 \nu_3  \cdots \nu_l ) } 
 + \cdots  ) + \cdots, \cr }}
where $a_1,a_2$ etc. are constants that can be worked out. 
The remaining terms multiplied by $a_2$ involve 
different contractions of one $ \mu $ index with one $\nu $ index.
The ``$\cdots$'' denote terms with more $\delta$'s which contract 
some set of $\mu $ indices with some set of $\nu $ indices 
multiplied by symmetric traceless tensors with lower rank. 
All the terms appearing on the right hand side of \gfprod\ are  
symmetric under the operation which exchanges all the $ \mu $ 
with the $\nu$ indices. This shows that the product is commutative
for an arbitrary pair of elements in $ \can ( S^4 )$. 

It is important to note that while the product 
$ A \circ B $ is commutative, it does not contain 
the most general element of $\chan( S^4)$  
appearing symmetrically  in the matrix product. 
\eqn\circnes{ A \circ B \ne { 1 \over 2 }  ( AB + B A ).  } 
We recall that the product $ A * B =  { 1 \over 2 }  ( AB + B A ) $
actually make $\chan( S^4)$  a Jordan Algebra, which requires
that the following identity be satisfied ( see  for example \gutze )  
\eqn\jord{ ( ( A*A) *B )* A = ( A * A ) * (B* A ). } 
The product $ A \circ B $ has an interesting relation 
to $ A * B $:
\eqn\circst{ 
A \circ B = A*B - Q( A B ). } 
$Q$ is a projector which acts as $1$ on all elements 
of  $\chan( S^4) $ which transform in representations 
of $ SO(5)$ corresponding to Young diagrams with rows 
$\vec r = ( 2L + M , 2L ) $ for $ L \ge 1 $ and $M$ is an 
arbitrary integer such that $ 2L + M \le n $. 
Equivalently, these Young diagrams have an even 
( non-zero ) number of columns of length $2$. 
$Q$ is zero on all other representations. 
Representations which are picked out by $Q$, 
while they involve some antisymmetrizations and are
projected out by $P$, have the property that they can 
appear symmetrically when $ A$ and $B$, belonging to 
the vector space $\can ( S^4)$, are multiplied 
as matrices. 

We observe that  $ Q ( X_{\mu} X_{\nu } ) = 0 $. Representations 
with an even number of columns of length two 
do not appear in the matrix product  $X_{\mu} X_{\nu }$, 
which means that the fuzzy sphere product coincides 
 with the symmetrized product and the identity in \jord\ 
is satisfied if $A$ and $B$ involve only $  X_{\mu}, X_{\nu}$. 
When we multiply more 
general elements in $ \can ( S^4 ) $ such as 
 $ X_{( \mu \nu ) } \circ X_{ ( \alpha \beta )} $
 we can get the representations picked out by $Q$. 

Representations which have $ \vec r = ( r_1, r_2 ) $
with $r_1, r_2 \ll n $, 
give harmonics which must be normalized by factors
$\co ( 1/n^{r_1} ) $, 
since they involve operators acting on $r_1$ different 
factors of the n-fold tensor product. Keeping track of such
normalizations 
\eqn\nrm{ 
{ X_{[k]}\over n^{k} }  { X_{[l] } \over n^l} 
 \sim { 1 \over n^{2} } { X_{[k+l-2, 2] } \over n^{k+l-2} }
+ \cdots, }
we have written down the simplest operator 
picked out by $Q$ which appears when symmetric traceless
tensor operators of rank $k$ and $l$ are multiplied
( the subscript on the $X$ describe the row lengths
of the irreducible representation they trasnform in ).  
If we only keep terms at order $1/n$ 
these can be neglected from the matrix product, 
so $ A \circ B = A *B $, which means that 
the Jordan identity is satisfied. 
Recall that the simplest representations which cause 
non-associativity 
of the $ A \circ B $ product are the 
representations with a second row of length $1$
 which can appear in the matrix product. This is a $1/n$ effect.

 To summarize, if we neglect all terms in the 
 structure constants of the normalized operators 
 which are $ \co ( 1/n) $ and higher, we have 
 a commutative and associative multiplication as 
 expected for an algebra approaching the classical 
 algebra of functions on the $S^4$. 
 If we keep $ \co ( 1/n) $ and neglect terms of 
 $ \co ( { 1\over n^2} )$ and higher, we have 
 commutative, non-associative algebra which is 
 an infinite dimensional Jordan algebra. 
 If we keep higher order terms we have a 
 commutative and non-associative algebra which is 
 not Jordan. And this is the structure at generic 
 finite $n$.

\listrefs

\end